\documentclass[aps,prx,reprint,twocolumn,superscriptaddress,floatfix,nofootinbib,longbibliography]{revtex4-1}
\usepackage{amsthm}
\usepackage{amsmath,amssymb,color,comment,physics}
\usepackage[makeroom]{cancel}
\usepackage[caption=false]{subfig}
\usepackage{mathrsfs}
\usepackage{graphicx}
\usepackage{float}
\usepackage{subfig}
\usepackage[countmax]{subfloat}
\usepackage[english]{babel}
\usepackage{dsfont}
\usepackage[bookmarks=true,colorlinks,linkcolor=RoyalBlue,urlcolor=NavyBlue,citecolor=RoyalBlue]{hyperref}
\usepackage[dvipsnames]{xcolor}
\usepackage{braket}
\usepackage{mathtools}
\usepackage{enumitem}
\usepackage{bm}
\usepackage{ragged2e}

\definecolor{mygreen}{rgb}{0.20,0.75,0.2}

\definecolor{myred}{rgb}{0.85,0.1,0.1}

\newcommand{\ee}[0]{\mathrm{e}}
\newcommand{\ii}[0]{\mathrm{i}}

\begin{document}
	
	\preprint{APS/123-QED}
	\title{Quantum many-body scars from unstable periodic orbits}

	\author{Bertrand Evrard}
	\email{bevrard@phys.ethz.ch}
	\affiliation{Institute for Quantum Electronics, ETH Z\"{u}rich, CH-8093 Z\"{u}rich, Switzerland}
	
	\author{Andrea Pizzi}
    \email{ap2076@cam.ac.uk}
	\affiliation{Department of Physics, Harvard University, Cambridge, Massachusetts 02138, USA}
    \affiliation{Cavendish Laboratory, University of Cambridge, Cambridge CB3 0HE, United Kingdom}
    
	\author{Simeon I. Mistakidis}
 \affiliation{Department of Physics, Missouri University of Science and Technology, Rolla, MO 65409, USA}
	\affiliation{ITAMP, Center for Astrophysics, Harvard $\&$ Smithsonian, Cambridge, Massachusetts 02138, USA}
	\affiliation{Department of Physics, Harvard University, Cambridge, Massachusetts 02138, USA}
	
	\author{Ceren B.~Dag}
	\email{ceren.dag@cfa.harvard.edu}
	\affiliation{ITAMP, Center for Astrophysics, Harvard $\&$ Smithsonian, Cambridge, Massachusetts 02138, USA}
	\affiliation{Department of Physics, Harvard University, Cambridge, Massachusetts 02138, USA}

\begin{abstract}
Unstable periodic orbits (UPOs) play a key role in the theory of chaos, constituting the ``skeleton'' of classical chaotic systems and ``scarring'' the eigenstates of the corresponding quantum system. Recently, nonthermal many-body eigenstates embedded in an otherwise thermal spectrum have been identified as a many-body generalization of quantum scars. The latter, however, are not clearly associated to a chaotic phase space, and the connection between the single- and many-body notions of quantum scars remains therefore incomplete. Here, we find the first quantum many-body scars originating from UPOs of a chaotic phase space. Remarkably, these states verify the eigenstate thermalization hypothesis, and we thus refer to them as \textit{thermal} quantum many-body scars. While they do not preclude thermalization, their spectral structure featuring approximately equispaced towers of states yields an anomalous oscillatory dynamics preceding thermalization for wavepackets initialized on an UPO. Remarkably, our model hosts both types of scars, thermal and nonthermal, and allows to study the crossover between the two. Our work illustrates the fundamental principle of classical-quantum correspondence in a many-body system, and its limitations.
\end{abstract}

\maketitle

\begin{figure*}
    \centering
    \includegraphics[width=\textwidth]{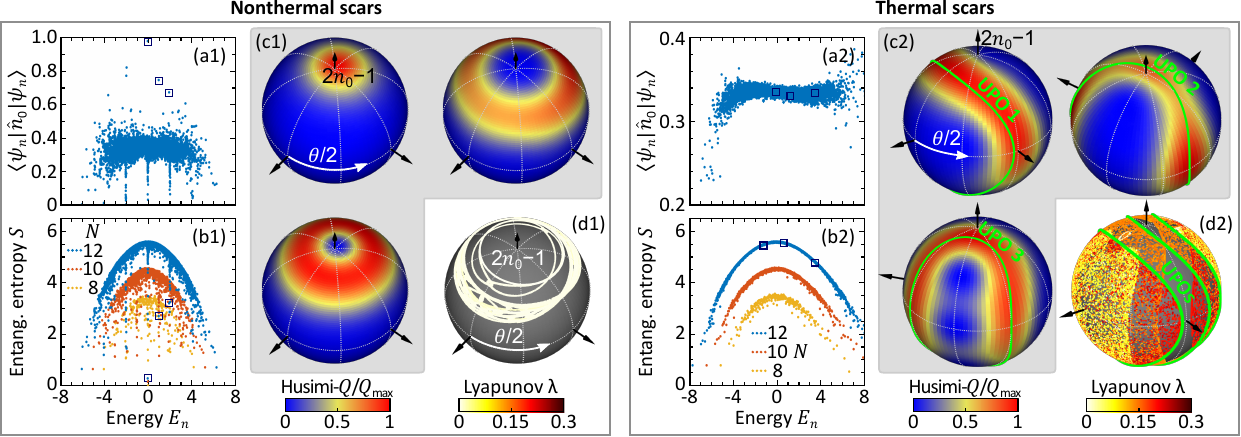}
    \caption{\textbf{Two types quantum of many-body scars.} Our model yields both standard nonthermal QMBS (left, a1-d1, for $p=0.05$) and thermal ones (right, a2-d2, for $p=0.5$). (a1,a2) The eigenstate expectation values of the population $\hat n_0$ obey weak (a1) and strong (a2) ETH. (b1,b2) In the middle of the spectrum, the half-chain entanglement entropy is maximal and follows volume-law scaling for most (b1) or all (b2) eigenstates. The Husimi distribution of the low entropy, ETH violating N-QMBS of the $p=0.05$ Hamiltonian in the $m=\eta=0$ plane are shown in panel (c1). These are  associated to regular orbits of the mean-field dynamics as shown in the Poincaré section (d1). For $p=0.5$, we observe quantum scars in the Husimi distribution (c2), which are instead associated to UPOs of the classical motion (d2). The selected eigenstates in (c1) and (c2) are marked by dark blue squares in (a1,b1) and (a2,b2), respectively. The Lyapunov exponents of the shown UPO's are $0.064$, $0.232$ and $0.328$.}
    \label{Figure1}
\end{figure*}

\section{Introduction}

The dichotomy between regularity and chaos is a fundamental feature of classical dynamical systems~\cite{strogatz2018nonlinear}. Regular regions of the phase space exhibit predictable patterns and stable orbits, while chaotic domains unfold into complex and diverging trajectories characterized by an extreme sensitivity to initial conditions. This intricate maze of chaotic trajectories is underpinned by a dense set of unstable periodic orbits (UPOs), which wield a pervasive influence on the entire dynamics~\cite{poincare1893methodes}, and constitute the ``skeleton'' of a chaotic phase space~\cite{cvitanovic2005chaos}.

Remarkably, UPOs play a central role not only in classical chaos, but also in quantum chaos. For a system with a semi-classical limit, a central question is how the structure of the classical phase space affects the spectral properties of the quantum counterpart. As for the energy spectrum, for instance, Gutzwiller's trace formula~\cite{Gutzwiller1971,gutzwiller2013chaos} expresses the density of states as a series expansion over the UPOs. As for the eigenstates, pioneering work by Heller~\cite{Heller1984,Heller1989} on chaotic quantum billiards has shown that the quantum wave function exhibits larger probability density around the UPOs of the underlying classical dynamics. In this sense, the classical UPOs \textit{scar} the quantum eigenstates.

The original and now 40 years old notion of quantum scars was developed for single-particle systems, e.g.,~quantum billiards. Motivated by recent developments in the coherent control of quantum many-body systems, and in particular by the experiments with trapped Rydberg atoms~\cite{Bernien2017}, a new notion of quantum many-body scars (QMBS) was proposed~\cite{Turner2018,turner2018quantum,moudgalya2018}. The QMBS are zero-measure and nonthermal eigenstates embedded in an otherwise thermal spectrum of states, featuring low entanglement entropy and giving rise to oscillatory dynamics for certain experimentally-accessible initial states~\cite{Bernien2017,PhysRevResearch.5.023010,zhang2023many}. Thanks to their ability to evade thermalization and break ergodicity, QMBS have attracted broad interest and now constitute an entire field of active research \cite{Serbyn2021,regnault2022quantum,Chandran_2023,Choi2019,Turner2018,Ho2019,Turner2021,moudgalya2018,moudgalya2018b,PhysRevB.101.220305,PhysRevB.102.195150,PhysRevB.102.241115,PhysRevLett.122.173401,PhysRevLett.123.147201,zhang2023many,PhysRevLett.126.210601,PhysRevB.107.L201105}. 

The names ``quantum scars'' and ``quantum many-body scars'' suggest a connection between the two. This connection is in fact far from granted. While both UPOs and QMBS are outliers in an otherwise chaotic system, QMBS are not associated with UPOs, as quantum scars by definition are. This ambiguity arises partly from the fact that most many-body systems lack a clear notion of classical phase space, in which the UPOs would reside. Recent works have addressed this challenge by constructing effective phase spaces based on variational ansatze~\cite{Ho2019,Michailidis2020,Turner2021}. In these systems, the quantum many-body state leaks out of the variational manifold into the rest of the Hilbert space, as quantified by a ``quantum leakage''~\cite{Ho2019,Michailidis2020}. The latter introduces the notion of instability inherent to quantum scars. At the same time, the leakage is not manifestly connected to an underlying classical instability, and indeed, in these works the QMBS correspond to regular regions of the variational phase space \cite{Michailidis2020}. To date, no example of a many-body system exists in which scars emerge in their original sense, namely in relation to UPOs in an underlying chaotic classical dynamics. 

Here, we report the first such example in a spin$-1$ chain, whose mean-field description provides a natural notion of a classical phase space. Therein, we identify stable and unstable periodic orbits, both scarring the eigenstates wave function to various degrees. The strongest form of scarring emerges from regular orbits with low quantum leakage. It consists of QMBS as described so far in the literature, i.e., weakly entangled and outlier eigenstates, violating the eigenstate thermalization hypothesis (ETH) \cite{Serbyn2021,regnault2022quantum}. We refer to such states as ``nonthermal QMBS''. By contrast in our model, UPOs underlie a weaker form of scarring, manifesting itself in ``thermal QMBS'', which are highly entangled and fulfill the ETH. Thermal QMBS do not preclude thermalization, but yield a particular spectral structure giving rise to an anomalous oscillatory dynamics when the system is initiated on an UPO. Upon varying a parameter of the model, we observe a crossover from nonthermal to thermal QMBS, which we analyse with respect to the stability of the underlying classical trajectory and the quantum leakage.

The remainder of this paper is structured as follows. In Sec.\,\ref{Sec: Two types of scars} we introduce the model, its mean-field description, and the two types of QMBS that it hosts. In Sec.\,\ref{Sec: N-QMBS} and \ref{Sec: T-QMBS} we describe the salient features of thermal and nontermal QMBS. In Sec.\,\ref{Sec: crossover} we discuss the crossover between the two types of scars. A summary of the results and an outlook on future perspectives are provided in Sec.\,\ref{Sec: Discussion}.

\section{Two types of quantum many-body scars}
\label{Sec: Two types of scars}

\subsection{Model}

We consider a one-dimensional spin$-1$ chain of size $N$ with periodic boundary conditions ($\hbar=1$),
\begin{align}
\hat H &= p\sum_{i=1}^N \hat s_{x,i} + \hat H_{\rm int}\,,\label{eq: Full Hamiltonian}\\
\hat H_{\rm int} &= \frac{1}{2}\sum_{i=1}^{N}\Big(\hat \kappa_{+,i}^\dagger\hat \kappa_{+,i+1}+\hat \kappa_{-,i}^\dagger\hat \kappa_{-,i+1}+
\mathrm{H.c.}\nonumber\\
&+\hat s_{z,i}\hat s_{z,i+1}\Big)\,,\label{Eq: Hint}
\end{align}
containing a transverse magnetic field of strength $p$ and a nearest neighbour interaction Hamiltonian $\hat H_{\rm int}$. Here $\bm{\hat s}_{i}$ are the spin-1 operators, and $\hat\kappa_{\pm,i}=\ket{0}_i\bra{\pm1}_i$, while $\mathrm{H.c.}$ stands for Hermitian conjugate. This Hamiltonian can be derived from a driven Heisenberg model, as shown in Appendix~\ref{App:derivation}.

\subsection{\label{ssec:MFdescription}Mean-field description}

First, we construct a classical description for our model by introducing coherent states $\vert \bm{\zeta}\rangle=\bigotimes_{i=1}^N\ket{\bm{\zeta}}_i\,$, i.e.,~product states with all sites in the same spin state $\ket{\bm{\zeta}}_i$. The complex vector 
$\bm{\zeta}$ is defined as
\begin{align}
	\bm{\zeta}=\begin{pmatrix}		\sqrt{n_+}\ee^{\phi_+}\\\sqrt{n_0}\ee^{\phi_0}\\ \sqrt{n_-}\ee^{\phi_-}
	\end{pmatrix}=\ee^{\phi_0}\begin{pmatrix}
	\sqrt{\frac{1-n_0+m}{2}}\ee^{\frac{\theta+\eta}{2}}\\\sqrt{n_0}\\ \sqrt{\frac{1-n_0-m}{2}}\ee^{\frac{\theta-\eta}{2}}
\end{pmatrix}\,,\label{Eq: MF spinor}
\end{align}
where we exploit the normalization condition $n_- + n_0 + n_+ = 1$, and introduce the magnetization $m=n_+-n_-$ and phases $\eta=\phi_+-\phi_-$ and $\theta=\phi_++\phi_--2\phi_0$. The global phase $\phi_0$ is irrelevant and hence set to zero. We can consider $n_0,m,\theta$, and $\eta$ as coordinates of a 4-dimensional classical phase space. Each point of such phase space is associated with one coherent state $\ket{\bm{\zeta}}$. The set of all the coherent states $\ket{\bm{\zeta}}$, which we call ``mean-field manifold'', serves as our variational manifold (Appendix~\ref{App: TDVP}). 

The equations of motion describing the dynamics projected onto this manifold are obtained using the time-dependent variational principle (TDVP)~\cite{Dirac1930,Haegeman_2011,Haegeman_2016,Ho2019,Michailidis2020}, see Appendix\,\ref{App: TDVP}, yielding~\cite{evrard2023quantum}
\begin{align}
	\dot n_0= & \hspace{1mm} p\sqrt{2n_0}\bigg [\sqrt{n_+}\sin \phi_+ + \sqrt{n_-}\sin\phi_- \bigg ]\,,\notag\\
	\dot \theta=& \hspace{1mm}2(1-2n_0) +p\bigg [\frac{2n_+-n_0}{\sqrt{2n_0n_+}}\cos\phi_+ \notag \\
	&+\frac{2n_--n_0}{\sqrt{2n_0n_-}}\cos\phi_-\bigg ]\,,\notag\\
	\dot m=& \hspace{1mm}p\sqrt{2n_0}\bigg [-\sqrt{n_+}\sin\phi_++\sqrt{n_-}\sin\phi_- \bigg]\,,\notag\\
	\dot\eta=& \hspace{1mm}-2m-p\sqrt{\frac{n_0}{2}}\bigg [\frac{\cos\phi_+}{\sqrt{n_+}}-\frac{\cos\phi_-}{\sqrt{n_-}}\bigg ]\,.\label{eq:eqnMotion}
\end{align}
The resulting dynamical system yields a rich mixed phase space with regular and chaotic regions~\cite{evrard2023quantum}. A key feature of this phase space is the existence of a family of orbits confined to the plane $m=\eta=0$ (for which indeed $\dot m=\dot\eta=0$ is straightforwardly verified). The Poincaré–Bendixson theorem~\cite{strogatz2018nonlinear} ensures that the dynamics on this two-dimensional manifold is periodic. However, there is a range of values of $p$ for which the plane $m=\eta=0$ intersects a chaotic region of the 4-dimensional phase space. In that situation, the periodic orbits are stable to in-plane perturbations and unstable to out-of plane ones, resulting in a positive Lyapunov exponent. The existence of UPOs is the precursor of quantum scarring~\cite{Heller1984}.

The projection of a quantum state onto the mean-field manifold can be visualized in the phase space via its Husimi distribution $Q(\bm{\zeta})=|\langle \bm{\zeta} | \psi\rangle|^2$. Note that $Q$ only depends on the projection of the many-body state $\ket{\psi}$ onto the exchange symmetric sector $\mathcal{H}_{\rm ex}$. The latter is defined as the space of all the states invariant under any permutation of the spins, of which the coherent states $\ket{\bm{\zeta}}$ form an overcomplete basis. Because the dimension of $\mathcal{H}_{\rm ex}$ only grows polynomially as~$N^2$, the von Neumann entanglement entropy of its states is at most logarithmic in $N$.

\subsection{Exact quantum treatment}

Two different regimes of our model, at $p=0.05$ and $p=0.5$, are highlighted in Fig.~\ref{Figure1} which compiles exact diagonalization \cite{QuSpin} and mean-field results. Our system possesses space-translation and spin-inversion symmetries, and we focus on the zero momentum and even parity symmetry sectors. Unless stated otherwise, we fix the system size to $N=12$. We present the expectation value of $\hat n_0 = \ket{0}_i\bra{0}_i$ for the energy eigenstates (independent of $i$ due to translational symmetry) in Figs.~\ref{Figure1}(a1) and (a2). For $p=0.5$, the expectation value $\bra{\psi_n} \hat n_0 \ket{\psi_n}$ is a smooth function of the energy $E_n$ up to fluctuations decreasing with system size, that is, the ETH is satisfied \cite{Deutsch1991,Rigol2008,DAlessio2016,Deutsch2018}. For $p=0.05$, instead, while most states conglomerate around similar values following the ETH, outliers exist (some marked with squares). Figs.~\ref{Figure1}(b1) and (b2) show the half-chain entanglement entropy $S=- \Tr{\hat\rho_{1/2} \log \hat\rho_{1/2}}$, where $\hat\rho_{1/2}$ is the reduced density matrix of half of the system, for each eigenstate $\ket{\psi_n}$. For both regimes, we find large entanglement entropy which is proportional to the subsystem volume as expected in chaotic many-body systems. For $p=0.05$, the outliers previously detected in panel (a1) also exhibit low entropy. Furthermore, their Husimi distributions illustrated in panel (d1) condense around stable quasi-periodic orbits of the regular mean-field dynamics, as can be seen in the Poincaré section provided in panel (c1).  We refer to such states as \textit{nonthermal} quantum many-body scars (N-QMBS). They match the usual properties of QMBS previously identified in the literature~\cite{turner2018quantum,Bernien2017,Serbyn2021,Chandran_2023}, and constitute the first instance of weak ergodicity breaking in the Hamiltonian of Eq.~\eqref{eq: Full Hamiltonian}.

\begin{figure*}
	\centering
	\includegraphics[width=\linewidth]{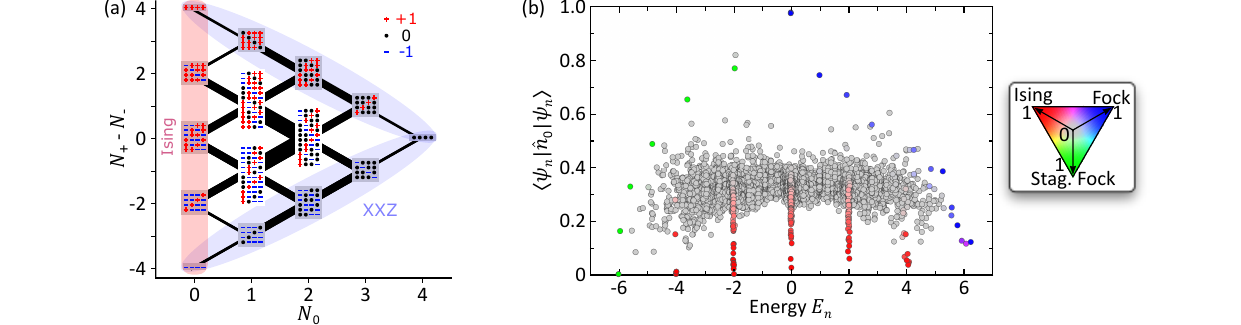}
	\caption{\textbf{Nonthermal QMBS} (a) Graph of the Hilbert space for a chain with $N=4$ sites. The states are grouped according to the population $N_0$ and magnetization $N_+-N_-$, which are good quantum numbers for the interaction Hamiltonian in Eq.~\eqref{Eq: Hint}. The links are proportional to the number of non-zero matrix elements of $S_x$ between the groups. For small enough $p$ (e.g.,~$p\approx 0.05$), the sectors located on the edge of the graph give rise to N-QMBS which can be described by effective spin-1/2 Hamiltonians, specifically the Ising and Heisenberg XXZ models. The latter has Fock-like eigenstates which provide an approximate description of N-QMBS states. Upon a simple staggering operation [see Eq.~\eqref{Eq: Staggering}] we obtain a family of ``staggered Fock states". In (b) we show that the nonthermal eigenstates of the full Hamilotnian in Eq.~\eqref{eq: Full Hamiltonian}, for $p=0.05$ and $N=12$, roughly correspond to these three sets of states: Ising, Fock, and staggered Fock. The colors in the eigenstate expectation value of $\hat n_0$ denote the overlap of the eigenstates at the chosen energy with the associated state emerging from the N-QMBS description (the color is obtained summing the three vectors in the colormap and weighting them by the respective projections).}
	\label{Figure2}
\end{figure*}

Turning now to $p=0.5$, we observe that the Husimi distributions of some special eigenstates exhibit enhanced probability around UPOs embedded in the chaotic sea, and have a positive Lyapunov exponent \cite{FootNotes_Lyapunov}. Unlike N-QMBS, such states satisfy ETH and have volume-law entanglement entropy; we thus refer to them as \textit{thermal} quantum many-body scars (T-QMBS). This new notion of T-QMBS might appear at first sight as an oxymoron. Indeed, thermal eigenstates are usually thought as random vectors~\cite{Deutsch1991,DAlessio2016,Deutsch2018}, having an ``ergodic" wave function and filling uniformly the available phase space from a semiclassical point of view~\cite{berry1984semiclassical}. The coexistence of thermal properties with ergodicity breaking in a variational manifold can be reconciled, providing that the variational manifold only captures a fraction of the eigenstate. This is indeed what we observe in our system where the weight of the eigenstates in the mean-field manifold remains small for T-QMBS (details in the App.\,\ref{App: T-QMBS}). Nevertheless, we show in the following that T-QMBS have a characteristic spectral structure which impact the quantum dynamics. Before delving in the in-depth description of T-QMBS, let us discuss the N-QMBS that exist in our model for low $p$.

\section{Nonthermal QMBS}\label{Sec: N-QMBS}

The N-QMBS of our model are best understood as eigenstates of the interaction Hamiltonian Eq.~\eqref{Eq: Hint} which shows robustness to small perturbations. The interaction Hamiltonian possesses $U(1)$ symmetries, namely $[\hat{H}_{\rm int},\hat N_m]=0$, where $\hat N_m = \sum_i \ket{m}_i\bra{m}_i$ counts the number of spins in state $m = 0,\pm 1$. In this model, N-QMBS emerge in three scenarios: (i) from the $N_0=0$ sector, (ii) from the $N_-=0$ sector (equivalent to $N_+=0$ owing to the spin inversion symmetry), and (iii) from the $N_+=N_-$ sector. In the following, we briefly describe these scenarios  (see Appendix\,\ref{App: N-QMBS} for more details). The results of this analysis are shown in Fig.~\,\ref{Figure2}, together with a graph highlighting the part of the Hilbert space most relevant to N-QMBS.

(i) \textit{$N_0=0$ sector}---The states in this sector are annihilated by the terms $\hat \kappa_{\pm,i}^\dagger\hat\kappa_{\pm,i+1}$, which can thus be omitted in $\hat H_{\rm int}$. The latter then reduces to the Ising model~\cite{sachdev1999quantum}, whose eigenstates form degenerate manifolds of states characterized by the number of domain walls $n_d$. The degeneracy of each manifold is $\begin{pmatrix} N\\n_d \end{pmatrix}$, and their energy is $E(n_d)=N/2-n_d$.

(ii) \textit{$N_-=0$ sector}---The states in this sector are annihilated by the terms $\hat \kappa_{-,i}^\dagger\hat\kappa_{-,i+1}$, which can thus be omitted in $\hat H_{\rm int}$.  On the other hand, the terms $\hat \kappa_{+,i}^\dagger$ and $\hat \kappa_{+,i}$ act as the raising and lowering operators for an effective spin$-\frac{1}{2}$ model, respectively. Within this representation, $\hat H_{\rm int}$ maps to the XXZ model, which can be solved with a Bethe ansatz \cite{sachdev1999quantum}. In particular, within each magnetization sector, the highest energy states are close to Fock states and provide a relatively good description of N-QMBS. This picture is consistent with the Husimi distribution of Fig.\,\ref{Figure1}\,(d1), condensed around a ring of constant $n_0$ with running $\theta = [0,4\pi)$. Furthermore, it provides intuition regarding the formation of N-QMBS. Indeed, Fock states have an exchange symmetry, which is preserved by $\hat S_x = \sum_i \hat s_{x,i}$, and are thus decoupled from all states that lack such symmetry.

(iii) \textit{$N_+=N_-$ sector}---For an even $N_+$, an additional set of N-QMBS can be obtained starting from those described in (ii). Let us consider a state where all spins are in the $m = +1$ state except some sites $i_1<i_2<...$ that are in the $m = 0$ state. For instance,
\begin{align}
	\ket{\psi(i_1,i_2,...)}=|+++ \underset{\mathclap{\substack{\uparrow \\ i_1}}}{0}++\,\underset{\mathclap{\substack{\uparrow \\ i_2}}}{0}\, ++  ...\rangle\,.
\end{align}
Let us now define a staggered counterpart of $\psi$
\begin{align}
	\ket{\psi_s(i_1,i_2,...)}=(-1)^{\sum_k i_k}|+-+ \underset{\mathclap{\substack{\uparrow \\ i_1}}}{0}-+\,\underset{\mathclap{\substack{\uparrow \\ i_2}}}{0}\, -+  ...\rangle\,.\label{Eq: Staggering}
\end{align}
We observe that the matrix elements of $\hat H_{\rm int}$ with respect to the states $\ket{\psi\{i_k\}}$ are exactly the opposite of those with respect to the states $\ket{\psi_s\{i_k\}}$, namely
\begin{equation}
\bra{\psi\{i_k\}} \hat H_{\rm int} \ket{\psi\{i_k^\prime\}}
=
- \bra{\psi_s\{i_k\}} \hat H_{\rm int} \ket{\psi_s \{i_k^\prime\}}.
\end{equation}
Therefore, from the N-QMBS of the $N_-=0$ sector with even $N_+$, which are built from the $\ket{\psi\{i_k\}}$ states, we obtain another set of N-QMBS with opposite energy.

We show in Fig.\,\ref{Figure2}(b) the expectation value of $\hat n_0$ for the eigenstates of the full spin$-1$ Hamiltonian \eqref{eq: Full Hamiltonian}. The colors reflect the overlaps of the eigenstates with the families of states that we identified as approximate descriptions of N-QMBS. Namely, we consider Fock states with $N_-=0$, their staggered counterpart, and the eigenstates of the Ising Hamiltonian (in order not to count the same state twice, we remove from the latter group the states $|+++...\rangle$ and $|+-+-...\rangle$, which are counted as Fock and staggered states, respectively). We observe that atypical states indeed have large overlaps with the states belonging to the three families of N-QMBS. The robustness of the symmetry sectors that we consider is qualitatively understood from their position on a Hilbert space graph, shown in Fig.\,\ref{Figure2}(a). These sectors are on the edges of the graph, and are weakly connected to the rest of the Hilbert space. Finally, we point out that a mean-field description is well suited for the Fock-like N-QMBS states as they belong to $\mathcal{H}_{\rm ex}$. The other two families of N-QMBS involve staggered configurations which would require an ansatz beyond mean-field, e.g.,~a matrix product state with small bond dimension. Hence we focus our attention in the rest of the paper on the N-QMBS in (ii) above.

The N-QMBS of our model are regularly spaced in energy, a feature that they share with many N-QMBS found in other models, and that is often related to the existence of an underlying spectrum generating algebra \cite{buvca2019non,Moudgalya2020etapairing,regnault2022quantum,Turner2018,Serbyn2021}. Such a structure can be straightforwardly obtained from our descriptions based on an effective spin-1/2 integrable Hamiltonian, i.e.,~in a similar fashion to \cite{evrard2023quantum}. It would be however only approximative for finite $p$. We provide details on the accuracy of these effective descriptions for N-QMBS in Appendix~\,\ref{App: N-QMBS}.

\section{Thermal QMBS}\label{Sec: T-QMBS}
\begin{figure}
    \centering
    \includegraphics[width=\linewidth]{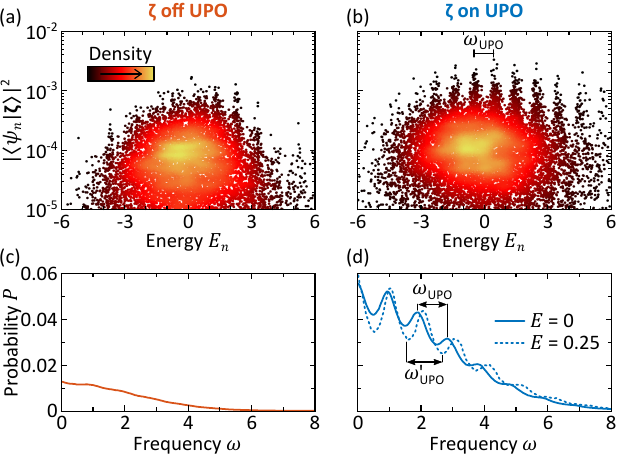}
    \caption{\textbf{Spectral structure of thermal QMBS} A hidden spectral structure of the T-QMBS emerges when projecting them on certain physically relevant wave functions. We show this by comparing coherent states $\ket{\bm{\zeta}}$ lying on or off an UPO. (a) When $\bm{\zeta}$ is off the UPO, the overlap of the coherent state with the eigenstates shows no unusual structure. (b) By contrast, if $\bm{\zeta}$ is on an UPO, a structure of approximately equispaced towers emerges. This structure is better captured by the weighted distribution $P(\omega)$ of the Bohr frequencies in Eq.\,(\ref{Eq: Bohr freq distrib}). (c) For $\bm{\zeta}$ off the UPO, the distribution $P(\omega)$ decays monotonously with frequency $\omega$. (d) For $\bm{\zeta}$ on the UPO, instead, $P(\omega)$ features peaks at the frequency $\omega_{\rm UPO}$ that characterizes the considered UPO, and at the multiples of $\omega_{\rm UPO}$. This characteristic frequency depends on the UPOs which the considered state $\bm{\zeta}$ overlaps with, as we show by considering a coherent state on a different UPO leading to a higher frequency $\omega_{\rm UPO}^\prime$. Note, in (a,b) the density of the markers is imprinted onto their color, whereas in (c,d) we have excluded the $\omega = 0$ component.}
    \label{Figure3}
\end{figure}

\subsection{Spectral structure of the scars}

We now turn to the study of the T-QMBS observed at $p=0.5$ in the Hamiltonian of Eq.~\eqref{eq: Full Hamiltonian}. These states cannot be distinguished from the rest of the spectrum by their entanglement entropy and expectation value of $\hat n_0$ (or other few-body observables). Yet, their Husimi distribution features a scar associated to the mean-field UPOs, suggesting that these eigenstates might be hiding an underlying structure. To uncover it, we investigate how the eigenstates overlap with the coherent states $\ket{\bm{\zeta}}$. More concretely, we choose two coherent states, one lying on the UPO and the other not. Furthermore, to diagnose the spectral structure of a given state $\ket{\psi}$ and identify the possible characteristic frequencies of the quantum dynamics from it, we introduce the following spectral density 
\begin{equation}\label{Eq: Bohr freq distrib}
P\left(\omega; \ket{\psi} \right) = \sum_{n<m}\delta(\omega-\omega_{nm})
 \left| \langle \psi | \psi_n \rangle 
 \langle \psi_m |\psi\rangle  \right|^2.
\end{equation}
This expression considers all possible Bohr frequencies $\omega_{nm}=E_m-E_n$ and weights them by the overlap of the involved eigenstates with $\ket{\psi}$. 

We first consider the case of a coherent state lying within the chaotic region, not on a UPO~\cite{FootNotes_NotOnUpo}, e.g.,~$\bm{\zeta}_{\rm chaos} \equiv\bm{\zeta}(n_0=1/2, \theta =0, m=0, \eta=4\pi/3)$. Note that the coherent states we consider are at zero energy, $\langle \hat H \rangle = 0$, and hence lie at the center of the spectrum. 
As expected for a Hamiltonian that satisfies ETH, the overlaps of the eigenstates with this excited coherent state has no particular structure, and the spectral density $P(\omega)$ is correspondingly featureless, exhibiting a simple and smooth decay with frequency $\omega$,
Fig.\,\ref{Figure3}(a,c).

The situation is markedly different for a coherent state still in the chaotic sea and at energy $\langle \hat H \rangle = 0$, but on the UPO, $\bm{\zeta}_{\rm UPO} \equiv \bm{\zeta}(n_0=1/2, \theta =4\pi/3, m=0, \eta=0)$. The eigenstates around zero energy are indeed all thermal, yet their overlap with $\ket{\bm{\zeta}_{\rm UPO}}$ highlights a remarkable underlying structure, as seen in Fig.~\ref{Figure3}(b). The overlaps arrange in towers approximately equally spaced in energy with a separation frequency corresponding to the UPO frequency $\omega_{\rm UPO}$. We note that our results are not contingent on the specific choice of $\bm{\zeta}_{\rm UPO}$. Indeed, the picture we drew here holds when varying $\bm{\zeta}_{\rm UPO}$ along the UPO, because the overlap with the coherent state is the Husimi function, $Q(\bm{\zeta}_{\rm UPO})$, which  is by definition higher along the UPO for T-QMBS, as  shown in Fig.\,\ref{Figure1}\,(c2). The spectral structure of the T-QMBS is best captured by the spectral density $P(\omega)$, depicted in Fig.\,\ref{Figure3}(d). $P(\omega)$ peaks approximately at the multiples of the UPO's characteristic frequency $\omega_{\rm UPO}$. We note that the frequency of the UPO depends on its energy. Correspondingly, we observe that a coherent state with 
$\bm{\zeta}_{\rm UPO}^\prime \equiv \bm{\zeta}(n_0=1/2, \theta, m=0, \eta=0)$ residing on another UPO at an energy $E=0.25$ with a larger frequency $\omega_{\rm UPO}^\prime$ accordingly yields a higher characteristic frequency.

\subsection{Unstable periodic orbits and many-body dynamics}
\begin{figure}
    \centering
    \includegraphics[width=\linewidth]{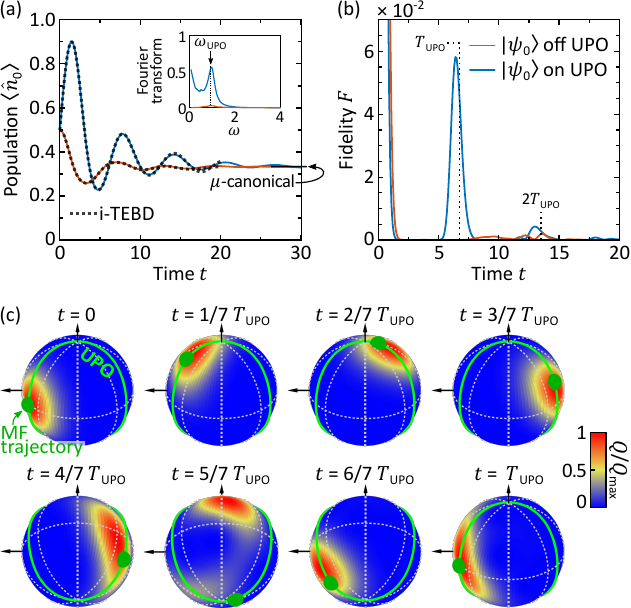}
    \caption{\textbf{Unstable periodic orbits and many-body dynamics}. The different spectral structure yielded by coherent states with $\bm{\zeta}$ on or off the UPO is reflected in the dynamics from these states. (a) In the former case, the expectation value of $\hat n_0$ features long-lived robust oscillations. The characteristic frequency of oscillation corresponds again to that of the underlying UPO, $\omega_{\rm UPO}$, as highlighted by the Fourier transform in the inset. The dashed line shows the i-TEBD results on quantum dynamics in the thermodynamic limit, and the microcanonical prediction, to which the system eventually relaxes, is marked. The solid lines are calculated with ED for $N=12$. (b) When initializing the system on a UPO, revival in the fidelity $F = \left| \langle \psi(0) | \psi(t) \rangle \right|^2$ is a characteristic of T-QMBS. (c) The effect of the UPO is further highlighted by means of the Husimi distribution, that evolves tracking the mean-field trajectory on the UPO marked in green. The green sphere on the UPO marks the instantaneous position of the classical trajectory. Calculations in (b) and (c) are done for a system size of $N=12$ with ED.}
    \label{Figure4}
\end{figure}

The spectral density in Eq.~(\ref{Eq: Bohr freq distrib}) for a state $\ket{\psi}$ is the Fourier transform of the time-evolved fidelity (up to a constant and a multiplicative factor),
\begin{equation}
\begin{aligned}
    F(t,\ket{\psi})&=|\bra{\psi}\ee^{-\ii\hat Ht}\ket{\psi}|^2\,\\
    &=\sum_{n,m}\ee^{-\ii (E_n-E_m)t} \left|
 \langle \psi | \psi_n \rangle 
 \langle \psi_m |\psi\rangle 
 \right|^2\,. \label{eq:survivalP}
\end{aligned}
\end{equation}
The striking difference in the spectral density for the states on and off the UPO thus implies qualitatively different dynamics for these states. Indeed, it is a general feature of quantum scars, both in the few- and many-body context, to result in fidelity revivals for special initial conditions~\cite{Pilatowsky_Cameo_2021,evrard2023quantum,Turner2018,Serbyn2021}.
In Fig.\,\ref{Figure4}, we first consider a coherent state 
\textit{off} the UPO, with $\bm{\zeta}_{\rm chaos}$, and observe the expected fast thermalization with no remarkable oscillatory behaviour in $\langle \hat n_0 \rangle$ and no fidelity revival. 

On the other hand, the dynamics starting from a coherent state \textit{on} the UPO, with $\bm{\zeta}_{\rm UPO}$, yields remarkable early-time oscillations in $\langle \hat n_0 \rangle$ and fidelity revivals at the frequency of the UPO. We further confirm this observation with the snapshots of the quantum dynamics at different times shown in Fig.\,\ref{Figure4}\,(c). The Husimi distribution of the many-body wave function at various times within the first period of the evolution shows how the ``wavepacket'' closely follows the UPO in phase space. The largest deviation from the UPO is seen when $n_0$ is minimum ($t\approx5/7 \ T_{\rm UPO}$ in Fig.\,\ref{Figure4}\,(c)) which can be understood from the fact that the leakage is maximal there, as discussed later in Sec.\,\ref{Sec: leakage}. Thermalization eventually occurs at long times because, in contrast to N-QMBS,  T-QMBS do obey the ETH. Let us note that a similar behavior is found for other few-body observables, e.g.,~$\hat n_+$ or $\hat s_x$, as well as for other coherent initial states taken on a UPO. Although it is difficult to simulate larger chains with $N>12$ due to the exponentially large Hilbert space, we test the robustness of the quantum dynamics via the infinite time-evolving block decimation (i-TEBD) simulations with a bond dimension of 250~\cite{tenpy}. Importantly, the early-time oscillations of $\langle \hat n_0\rangle$ persist in the thermodynamic limit $N \to \infty$, see the dashed line in Fig.\,\ref{Figure4}\,(a). In that limit the total fidelity vanishes, but we have verified with i-TEBD simulations that the revivals of the fidelity per site~\cite{Michailidis2020} are robust.

\section{From nontermal to thermal QMBS}
\label{Sec: crossover}

\subsection{Periodic orbits at $E=0$: a pendulum motion}
Having shown the existence of two types of QMBS for $p = 0.05$ and $p = 0.5$, we now describe the crossover between the two as $p$ is varied.
\begin{figure}
	\centering
	\includegraphics[width=\linewidth]{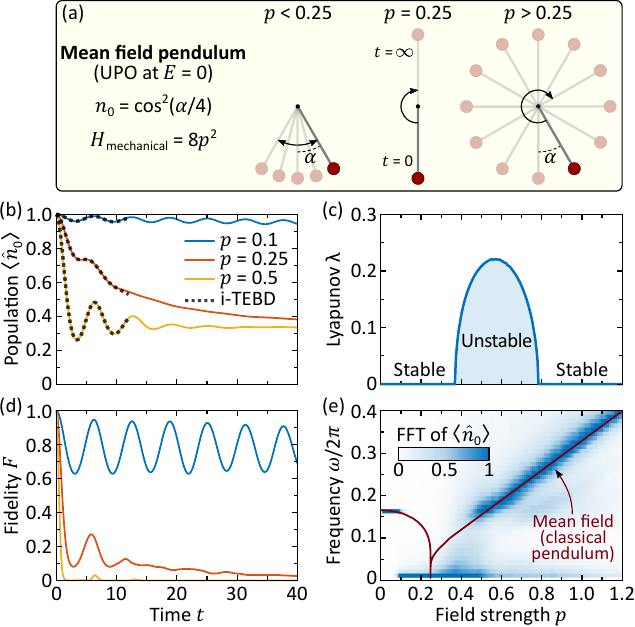}
	\caption{\textbf{Periodic orbits at $E=0$: a pendulum motion} (a) Pendulum picture for the orbits with energy $E=0$ in the plane $m=0,\eta=0$. The initial state with $n_0 = 1$ corresponds to the pendulum with position $\alpha = 0$ and kinetic energy $8p^2$. Three dynamical behaviors are possible: oscillatory ($p<0.25$), homoclinic ($p=0.25$), and rotating ($p>0.25$). The regimes of the classical pendulum are reflected in the quantum dynamics at short time, both for the observable $\langle \hat n_0 \rangle$ (b) and for the fidelity $F$ (d). At longer time, we observe a crossover from oscillating ($p\lesssim0.25$) to thermalizing ($p\gtrsim0.25$) behaviors which is understood from the extension of the trajectory toward the phase space region with high quantum leakage (small $n_0$, see Fig.~\ref{Figure6}).  (e) The Fourier transform of $\langle\hat n_0\rangle$ shows excellent agreement with the classical pendulum analytic prediction, Eq.~\eqref{Eq: Pendulum frequency}. For each value of $p$ the Fourier transform is normalized to its maxima (excluding the zero frequency). Remarkably, the signature of the pendulum orbits is observed even when the Lyapunov exponent takes on a finite value (c), that is, when the classical trajectory becomes a UPO.}
	\label{Figure5}
\end{figure}
We focus on a coherent state with $\bm{\zeta}_0 = (0,1,0)$, that is, with all the spins in the $m=0$ state. This state has energy $\langle \hat H \rangle = 0$ independently of $p$, and thus lies in the center and most chaotic region of the spectrum. Conveniently, $\ket{\bm{\zeta}_0}$ is an eigenstate of the interaction Hamiltonian $\hat H_{\rm int}$, from which N-QMBS emerge at small $p$.  Moreover, $\ket{\bm{\zeta}_0}$ lies on a periodic orbit of the $m=\eta=0$ manifold, which turns from stable to unstable for increasing $p$, as seen in Fig.\,\ref{Figure5}(c). Thus, the state $\ket{\bm{\zeta}_0}$ is ideal to study the crossover from N- to T-QMBS as $p$ is varied.

We begin by describing the mean-field dynamics. Using the conservation of energy, and the change of variable $\alpha=4\arccos(\sqrt{n_0})$, we obtain an orbit equation (see Appendix\,\ref{App: Pendulum})
\begin{align}
	\frac{\dot\alpha^2}{2}+\frac{1}{4}(1-\cos\alpha)=8p^2\,.\label{Eq: Pendulum}
\end{align}
This equation yields the mechanical energy of an effective pendulum with unit momentum of inertia, length, and gravitational acceleration, and with mechanical energy 
$8p^2$, see Fig.\,\ref{Figure5}(a).  The initial condition $n_{0}(0)=1$ corresponds to an angle $\alpha(0)=0$ and an angular velocity $\dot\alpha(0)=4p$. For $p<1/4$\,, the pendulum undergoes oscillatory dynamics with $\alpha\in(-\pi,\pi)$ and $n_0\in(1/2,1]$. For $p>1/4$, instead, the large energy allows the pendulum to rotate indefinitely with $\alpha$ growing unbounded, and $n_0$ undergoing full oscillations between $0$ and $1$. The value $p = 1/4$ is highly singular: not only the period of the oscillations diverges, i.e.,~the pendulum reaches the upright position at infinite time, but also the oscillation amplitude of $n_0$, that depends on $\alpha$ modulo $[4\pi]$, has a discontinuity. The frequency of oscillations of $n_0$ reads~\cite{landau_lifshitz_vol1}
\begin{equation}
\omega_0 =
\begin{cases}
    \dfrac{\pi}{2K(4p)} & \mathrm{for} \quad p<\dfrac{1}{4} \\[12pt]
    \dfrac{\pi p}{K(1/4p)}
    & \mathrm{for} \quad p>\dfrac{1}{4},
\end{cases}
\label{Eq: Pendulum frequency}
\end{equation}
where $K$ denotes the complete elliptic integral of the first kind. We compare these predictions with the quantum dynamics in Fig.\,\ref{Figure5}, and find that 
the short time behavior is well captured by the pendulum dynamics as illustrated in panel (b). Namely, for $p=0.1$ there are oscillations with small amplitude; at $p=0.25$, the pendulum is on a homoclinic orbit where the period diverges and no oscillations occur; for $p=0.5$ we recover the oscillations with a larger amplitude and a faster damping. The frequencies of the oscillations in $n_0$ are in good agreement with the mean-field pendulum prediction, as shown in panel (e). 
Consistently, we observe persistent revivals in the fidelity in  Fig.\,\ref{Figure5}(d) for $p=0.1$ as expected for N-QMBS, a slow fidelity decay for $p=0.25$ and rather small revivals for T-QMBS at $p=0.5$.

\subsection{Quantum leakage}\label{Sec: leakage}
\begin{figure}
    \centering
    \includegraphics[width=\linewidth]{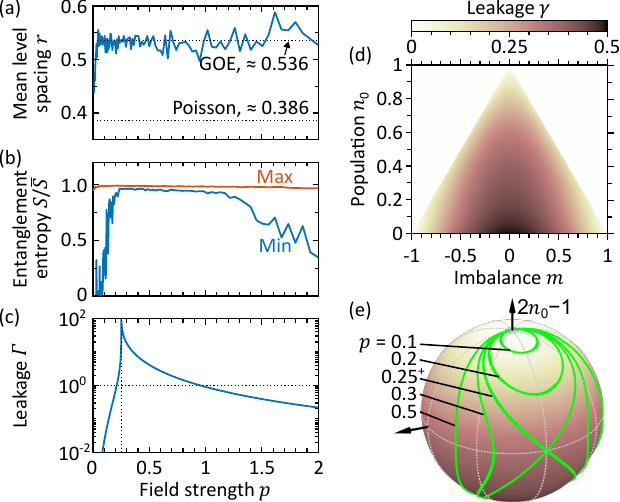}
    \caption{\textbf{From nonthermal to thermal QMBS.} (a) Mean level spacing ratio $r$, with Poisson and GOE reference values ($2 \log 2 - 1 \approx 0.386$ and $4-2\sqrt{3} \approx 0.536$, respectively~\cite{atas2013distribution}). The GOE value is reached already for a small $p$, indicating quantum chaos. 
    (b) A stringent diagnostic of thermalization is provided by the maximum and minimum half-chain entanglement entropy computed over a window of 400 eigenstates around $E=0$. For $0.25 \lessapprox p 	\lessapprox 1.2$, all the eigenstates approximately saturate the maximal entropy $\bar{S} = \log \mathcal{D}$, with $\mathcal{D}$ the dimension of the Hilbert space.
    (c) Quantum leakage for the periodic trajectory initiated at $n_0=1$. The leakage diverges at $p=0.25$ which relates to the pendulum dynamics shown in Fig.~\ref{Figure5}. We observe that the crossover from nonthermal- to thermal-QMBS ($p\sim 0.25$) and back ($p\sim1.2$), evidenced by the disappearance and resurgence of weakly entangled state in (b), occurs when the quantum leakage rises above $\Gamma\sim1$ observed in (c). (d) Instantaneous quantum leakage $\gamma$. The latter only depends on $n_0$ and $m$, see Eq.~\eqref{Eq. leakage explicit}, and can thus be represented on a plane. (e) Instantaneous quantum leakage on the $m = \eta = 0$ manifold, where it only depends on $n_0$. UPOs at energy $E = 0$ are displayed for various $p$. The UPOs cross the high-leakage region (southern hemisphere) for $p>0.25$.}
    \label{Figure6}
\end{figure}
The mean-field picture outlined above, expressed in terms of the effective pendulum at $E = 0$, captures some key aspects of the early dynamics. However, it fails to predict the observed thermalization for $p\gtrapprox0.25$ in the quantum dynamics. Indeed, the wave function does not remain on the mean-field variational manifold and leaks into the rest of the Hilbert space due to growing entanglement. This phenomenon is named quantum leakage~\cite{Ho2019,Michailidis2020,Haegeman_2011,Haegeman_2016}, and its rate is denoted by $\gamma$. The latter is defined on each point $\bm{\zeta}$ of the variational manifold as the norm of the difference between the exact time derivative of the mean-field state $\ket{\bm{\zeta}}$ given by the Schr\"odinger equation, and the mean-field derivative obtained from the TDVP \cite{FootNotes_NormalizationGamma}, namely
\begin{align}
	\gamma(\bm{\zeta}) = \frac{1}{\sqrt{N}}\left\Vert \ii\hat H\ket{\bm{\zeta}}+\left(\frac{d}{dt}\ket{\bm{\zeta}}\right)_{\rm mf}\right\Vert\,.\label{Eq: def leakage}
\end{align}
We point out that the single-particle term $p\sum_i\hat s_{x,i}$ in Eq.~\eqref{eq: Full Hamiltonian} yields the same contribution to exact and mean-field derivatives. Hence, only the interaction Hamiltonian $\hat H_{\rm int}$ contributes to the leakage, which is thus independent of $p$. Furthermore, the $U(1)$ symmetries of $\hat H_{\rm int}$ imply that the leakage only depends on the population $n_0$ and magnetization $m$, and not on the phases $\theta$ and $\eta$. By an explicit calculation, detailed in Appendix~\ref{App: TDVP}, we find
\begin{align}
\gamma(\bm{\zeta}) &=\left[\frac{1}{4}\left(n_0(1-2n_0)+m^2-1\right)^2-n_0^3\right]^{1/2}\,.
\label{Eq. leakage explicit}
\end{align}
The leakage is presented in Fig.~\ref{Figure6}(d) as a function of $n_0$ and $m$. It vanishes for $(n_0,m)=(1,0)$ and $(n_0,m) = (0,\pm 1)$ corresponding to three eigenstates of $\hat H_{\rm int}$. Following Refs.~\cite{Ho2019,Michailidis2020}, for a given periodic orbit with period $T$ we define a dimensionless and time-averaged leakage
\begin{align}
    \Gamma=\left[\int_{0}^{T}dt\gamma(\bm{\zeta}(t))\right]^2\,,\label{Eq: Integrated leakage}
\end{align}
where $\bm{\zeta}(t)$ is the solution of the mean-field equations of motion along the periodic orbit. Qualitatively, $\Gamma^{-1}$ can be understood as a quality factor for the trajectory: a large $\Gamma$ signals a fast leak of the state from the mean-field manifold, leading to thermalization and a quick breakdown of the mean-field description.

Note that while $\gamma$ does not depend on $p$, the trajectory $\bm{\zeta}(t)$ and hence $\Gamma$ do, as shown in Fig.\,\ref{Figure6}\,(c) for the ``pendulum orbit" with $n_0=1$. In the limit $p\to0$, $n_0=1$ is a stationary point with vanishing leakage $\gamma(n_0=1)=0$ such that $\Gamma\to0$. In the oscillating regime of the pendulum, e.g.,~$0<p<1/4$, the amplitude and period of oscillation increase with $p$, letting the system to spend more time in the high leakage region of the phase space [see Fig.\,\ref{Figure6}\,(d,e)] and resulting in an increase of $\Gamma$. At $p=1/4$, the period of the periodic orbit and hence $\Gamma$ diverge. Increasing $p$ further, the period decreases as $T\sim 1/p$ [Fig.\,\ref{Figure5}\,(e)], and so does $\Gamma$. This analysis explains the damping of the oscillations observed in Fig.\,\ref{Figure5}(b) and (d). We point out that while the damping is correlated with the quantum leakage, it appears to be insensitive to the value of the Lyapunov exponent. In particular, for $p\approx0.8$, the classical motion turns regular, Fig.~\ref{Figure5}(c), with no sharp effect on the quantum dynamics. Here the quantum leakage is the fastest timescale, and the system leaves the mean-field manifold before the classical chaotic dynamics can be realized.

Let us now relate the leakage with the presence of N- and T-QMBS. In Fig.\,\ref{Figure6}\,(a), we show the mean nearest-neighbor level spacing ratio $r$ which quickly rises from $r\approx0.38$ at $p=0$, to $r\approx0.53$ at $p\approx0.03$, matching the predictions of Poisson and Gaussian Orthogonal Ensemble (GOE) distributions \cite{atas2013distribution}, respectively. This indicates a rapid breakdown of integrability with $p$. The presence of N-QMBS is witnessed by the difference between the minimum and maximum entanglement entropy, $S_{\rm max}-S_{\rm min}$, shown in panel (b) with respect to $p$. Indeed, this occurs for $0.03 \lessapprox p \lessapprox 0.25$, when the spectrum is mostly chaotic ($r \approx 0.53$), but N-QMBS have low entropy and are responsible for finite $S_{\rm max}-S_{\rm min}$. For $p\gtrapprox0.25$ all N-QMBS disappear, and the system reaches a fully thermal regime with all states featuring volume-law entanglement. For $p\gtrapprox1.2$, the interaction progressively becomes negligible, integrability is gradually restored and nonthermal states re-emerge in the spectrum. These observations are consistent with the behavior of the quantum leakage provided in panel (c) and with the conjecture that N-QMBS should exist for $\Gamma\lesssim1$~\cite{Michailidis2020,Ho2019}.

\section{Discussion and Outlook}\label{Sec: Discussion}

To summarize, we have observed a new manifestation of quantum scarring in a many-body system. In the past, many-body scars have been associated to regular variational phase spaces. Instead in our work, we show many-body scars that are underpinned by UPOs in a chaotic phase space. The UPOs affect the many-body eigenstates, and consequently the quantum dynamics. The existence and properties of QMBS associated with classical trajectories result from the interplay between classical and quantum instabilities. The former are related to the divergence of nearby classical trajectories within the variational manifold, as measured by the Lyapunov exponent $\lambda$. The latter originate from the breakdown of the variational approach, and are quantified by the quantum leakage $\gamma$,  or more precisely by its average along a periodic orbit $\Gamma$, Eq.~\eqref{Eq: Integrated leakage}. Until now, two scenarios were known where either a classical or a quantum instability was present. Our work highlights yet a richer landscape of many-body scarring, in which the condition $\Gamma \lessapprox 1$ can coexist with a finite $\lambda$, and the many-body scars are underpinned by UPOs. Let us briefly review the salient features of these regimes.

Scenario (i) -- $\lambda>0$ and $\Gamma\to0$, that is, chaotic classical dynamics which becomes exact in a semi-classical limit. This corresponds to the original notion of quantum scars as encountered in quantum billiards. In this context, scars emerge if $\Lambda=\lambda T/(2\pi) \lessapprox 1$~\cite{Heller1984} which explains why, among the infinitely many UPOs, only those with short period can result in scarring. This scenario can also be met in a many-body model featuring all-to-all interactions, where the mean-field limit is exact for $N\to\infty$. Refs.~\cite{evrard2023quantum,Pilatowsky_Cameo_2021} found that quantum scars are ubiquitous in this case, and do not necessarily compromise thermalization.

Scenario (ii) -- $\Lambda=0$ and $\Gamma\ll 1$, that is integrable classical dynamics and low quantum leakage. This underpins nonthermal eigenstates, long-lived oscillatory quantum dynamics, and lack of thermalization. This phenomenolgy,   motivates the name \textit{nonthermal} QMBS~\cite{Michailidis2020}. The notion of instability in this case stems from the presence of strong quantum leakage in the rest of the phase space. This situation is met in our model for $p\lesssim0.25$, and was previously encountered in the PXP model \cite{Ho2019,Michailidis2020,Turner2021}.
    
Scenario (iii) -- $\Lambda\sim1$ and $\Gamma\sim1$, that is, chaotic classical dynamics and relatively large quantum leakage. This scenario is achieved in our model for $p\sim[0.4,0.8]$ where the signatures of the classical orbits on the eigenstates are weaker. In particular, all states verify the ETH, and are highly entangled. Nevertheless, we uncover towers of \textit{thermal} QMBS, scarred by the UPOs. As a result, the early-time quantum dynamics follows the classical trajectory. The latter is embedded in a chaotic region, which strengthens the analogy with the original single-particle quantum scars. The quantum instability dominates over the classical one and the system leaves the variational manifold before it is impacted by the classical instability. As a consequence,  quantum dynamics is insensitive to variations of the Lyapunov exponent, Fig.\,\ref{Figure5} (c,e).

In our model, the onset of the classical and quantum instability are concomitant and we observe a fast crossover from (ii) to (iii) as $p$ is increased. This is intuitively understood from the fact that as the classical dynamics turns ergodic, it becomes harder to find an orbit confined in the small phase space region where the leakage is low. While this qualitative reasoning might suggest a direct relation between $\Lambda$ and $\Gamma$, this is not granted. For instance, changing the range of the interaction (with a proper rescaling of its strength) does not modify the mean-field dynamics, but it alters $\Gamma$. Based on these observations, we can envision a fourth scenario:

Scenario (iv) -- $\Lambda\sim1$ and $\Gamma\ll1$, that is, a UPO confined within the low leakage region of a chaotic phase space. It is an open question whether these conditions would result in thermal eigenstates. Indeed, even for all-to-all interactions, where $\Gamma\to0$ in the thermodynamic limit (see Appendix \,\ref{App: TDVP}), the ETH was found to be verified \cite{evrard2023quantum}. These conditions remain to be seen, and designing a Hamiltonian~\cite{Choi2019,Michailidis2020_Stabilizing,Hallam2023} that realizes this regime is an exciting prospect for future research.

Our work paves the way towards a tentative classification of QMBS based on the Lyapunov exponent and the quantum leakage together.  Indeed, chaotic variational phase spaces are ubiquitous in many-body systems~\cite{Hallam2019}. Hence it is imperative to extend our semi-classical analysis of QMBS 
by employing different models and generalizing our approach. Recently, considerable work has been devoted to the study of quantum scars in many-body systems with a semiclassical limit (e.g., with all-to-all interactions)~\cite{sinha2023classical,sinha2020chaos,Sinha2021,vardi2010bjj, khripkov2013coherence, Pilatowsky_Cameo_2021, Pilatowsky-Cameo_2021_NJP,PhysRevLett.130.250402,evrard2023quantum}. Such systems possess a permutation symmetry and are thus amendable to a mean-field description which becomes exact in the thermodynamic limit. Thereby, a classical analysis and a connection to single-particle scars is straightforward. In particular, the all-to-all limit of our model was studied in Ref.~\cite{evrard2023quantum}. Starting from such collective models, one can then break the permutation symmetry and thus the semiclassical solvability either by reducing the range of the interaction~\cite{lerose2023theory}, or by introducing disorder. In this way, one has a knob to continuously steer the system away from its classical limit, while tracking the fate of the quantum leakage and of the scars.

\begin{acknowledgments}
We thank Anushya Chandran, Alexey Khudorozhkov, Maksym Serbyn, Luca Tagliacozzo, Amichay~Vardi, and Norman Y.~Yao for helpful comments and discussions. B.~E. is supported by an ETH Zurich Postdoctoral Fellowship. A.~P. is supported by the AFOSR MURI program (Grant No. FA9550-21-1-0069). C.~B.~D. and S.~I.~M. acknowledge financial support from the NSF through a grant for ITAMP (Award No: 2116679) at Harvard University. This research was supported in part by grants NSF PHY-1748958 and PHY-2309135 to the Kavli Institute for Theoretical Physics (KITP). 
\end{acknowledgments}
 
\appendix

\section{Derivation of Eq.~\eqref{eq: Full Hamiltonian}\label{App:derivation}}

We start from the following periodically driven Heisenberg Hamiltonian, 
\begin{eqnarray}
\mathcal{\hat H}(t) &=& \sum_{i=1}^{N} \bm{\hat s_{i}} \cdot \bm{\hat s_{i+1}} - q \hat s_{z,i}^2+ p \cos (qt) \hat s_{x,i}\,,\label{eq:initHamiltonian}
\end{eqnarray}
where $\hat s_{x,y,z,i}$ correspond to spin$-1$ operators at site $i$. We apply a unitary transformation $\hat U =\prod_i \exp\left(-\ii qt\hat s_{z,i}^2\right)$ to Eq.~\eqref{eq:initHamiltonian}. It is convenient to express $\hat s_{x,y}$ in terms of $\hat\kappa_{\pm,i}=\ket{0}_i\bra{\pm1}_i$ as
\begin{align}
    \hat s_x &= \frac{1}{\sqrt{2}}(\hat\kappa_++\hat\kappa_-)+\mathrm{H.c}\,\\
    \hat s_y &= \frac{1}{\sqrt{2}\ii}(\hat\kappa_++\hat\kappa_-)+\mathrm{H.c}\,,
\end{align}
and the $\hat\kappa_{\pm,i}$ operators transform under $U$ as in
\begin{align}
    \hat U^\dagger\hat\kappa_{\pm,i}\hat U = \ee^{\ii\mp qt}\hat\kappa_{\pm,i}\,.
\end{align}
This makes transforming the spin operators under $U$ straightforward. We obtain
\begin{eqnarray}
    \hat U^{\dagger} \hat s_{x,i} \hat U &=& \frac{1}{\sqrt{2}} \big(  e^{iqt} \hat \kappa_{-,i} + e^{-iqt} \hat \kappa_{+,i} \notag \\
    &+&  e^{iqt} \hat \kappa^{\dagger}_{+,i} + e^{-iqt} \hat \kappa^{\dagger}_{-,i}\big). \notag
\end{eqnarray}
Defining the time-averaging operation as $\mathcal{T}[f]=1/T\int_0^T f(t) dt$ where $T=\frac{2\pi}{q}$ is the period of the drive, for $q\gg1$ we can neglect the fast rotating terms, namely
\begin{align}
p \cos (qt) \hat U^{\dagger} \hat s_{x,i} \hat U
& \approx
 \mathcal{T} [p \cos (qt) \hat U^{\dagger} \hat s_{x,i} \hat U]\\
 & = \frac{p}{\sqrt{2}} \big(  \hat \kappa_{-,i} + \hat \kappa^{\dagger}_{-,i} + \hat \kappa_{+,i} +  \hat \kappa^{\dagger}_{+,i}  \big) \notag \\
 & = p \hat s_{x,i}.
\end{align}
With analogous considerations we obtain
\begin{align}
\hat U^{\dagger}  \bigg(\hat s_{x,i} \hat s_{x,i+1} + & \hat s_{y,i} \hat s_{y,i+1}\bigg)\hat U \notag \\
& \approx \hat  \kappa_{-,i}^{\dagger}\hat \kappa_{-,i+1}  + \hat \kappa_{+,i}^{\dagger}\hat \kappa_{+,i+1} + \textrm{h.c.}
\end{align}
The term $\hat s_{z,i} \hat s_{z,i+1}$ remains invariant under the transformation. This leads to Eq.~\eqref{eq: Full Hamiltonian}. We note that the mean-field limit of Eq.~\eqref{eq:initHamiltonian} is provided by all-to-all interacting model in Ref.~\cite{evrard2023quantum}.

\section{Nonthermal QMBS}\label{App: N-QMBS}
\begin{figure}
    \centering
    \includegraphics[width=\linewidth]{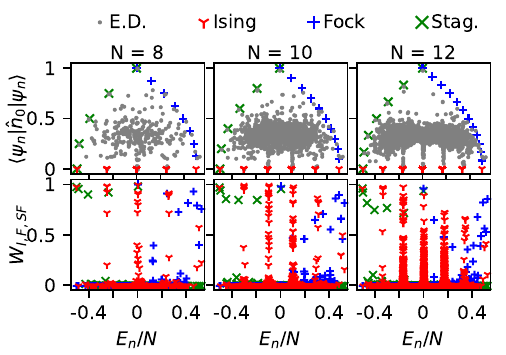}
    \caption{The upper panels show the eigenstate expectation values of $\hat n_0$ obtained from exact diagonalization of the full Hamiltonian (grey dots). We look at different system sizes $N$ and compare the results with the prediction of the effective representations, introduced in Sec.\,\ref{Sec: N-QMBS} and further discussed in Appendix \,\ref{App: N-QMBS}. 
    The lower panels show the weight of the eigenstates within the subspaces spanned by the three families of states introduced in the main text.}
    \label{FigureSM_RQMBS}
\end{figure}

In this appendix, we provide more details on the description of the N-QMBS. As mentioned in Sec.\,\ref{Sec: N-QMBS}, we identified three families of N-QMBS which are close to eigenstates of the interaction Hamiltonian $\hat H_{\rm int}$. The first two families emerge from the $N_0=0$ and $N_\pm=0$ symmetry sectors of $\hat H_{\rm int}$. Due to the absence of one of the spin states ($m=0$ or $m=+1$ or $m=-1$), $\hat H_{\rm int}$ can be mapped onto an effective spin-1/2 model. For the $N_0=0$ sector, the spin-flip terms $\kappa_{\pm,i}^\dagger\kappa_{\pm,i}$ vanish and the Ising model is obtained.

For the $N_-=0$ sector (or equivalently for $N_+=0$), we introduce effective spin-1/2 operators:
\begin{align}
    \hat j_x &= \frac{1}{2}(\ket{0}\bra{+}+\ket{+}\bra{0})=\frac{1}{2}(\hat \kappa_++\hat \kappa_+^\dagger)\,,\nonumber\\
    \hat j_y &= \frac{\ii}{2}(\ket{0}\bra{+}-\ket{+}\bra{0})=\frac{\ii}{2}(\hat \kappa_+-\hat \kappa_+^\dagger)\,,\nonumber\\
    \hat j_z &= \frac{1}{2}(\ket{+}\bra{+}-\ket{0}\bra{0})\,.
\end{align}
Note that $[\hat \kappa_+^{\dagger},\hat \kappa_+]=2\hat j_z$ and $[\hat j_x,\hat j_y]=i \hat j_z$ following the  SU(2) algebra. $H_{\rm int}$ projected on the sector with $N_-=0$, maps onto the integrable XXZ chain when written in terms of $\bm{\hat j}$ operators,
\begin{align}
    \hat H_{\rm XXZ} \sim \sum_i&\left[\hat j_{x,i}\hat j_{x,i+1}+ \hat j_{y,i} \hat j_{y,i+1}+\frac{1}{2}(\hat j_{z,i}\hat j_{z,i+1}+\hat j_{z,i})\right]\,,\label{Eq: XXZ Hamiltonian}
\end{align}
up to a constant $L/8$. It is seen that the magnetization $\hat J_z=\sum_i \hat j_{z,i}$ is conserved, i.e.,~$[\hat H_{\rm XXZ},\hat J_z]=0$. Within each magnetization sector, the highest energy eigenstate appears to be relatively robust to the perturbation $p\hat S_x$ giving rise to N-QMBS. This can be understood within the Bethe ansatz where the highest energy states are those with minimum momenta (for antiferromagnetic interaction), and are thus close to Fock states~\cite{gaudin2014bethe}. Furthermore, Fock states are invariant upon any permutation of the sites since they belong to $\mathcal{H}_{\rm ex}$, and this symmetry is preserved by the action of $\hat S_x$. Therefore, they are decoupled from the rest of the Hilbert space which qualitatively explains their special robustness to the $p\hat S_x$ perturbation. We point out that upon a slight change of the model, namely by multiplying the $s_{z,i}s_{z,i+1}$ interaction by two, $\hat H_{\rm int}$ is mapped onto the isotropic Heisenberg chain where the Fock states are exact eigenstates. However, we did not observe that this change of the Hamiltonian results in a qualitative change regarding the nature of the N-QMBS or their robustness against the $p\hat S_x$ perturbation.

As mentioned in the main text, a third family of N-QMBS can be obtained from the one described above by replacing the ``ferromagnetic" sequence such as $|++0++..\rangle$ by the staggered ``antiferromagnetic sequence" $|+-0+-..\rangle$. This flips the sign of the interaction $s_{z,i}s_{z,i+1}$ in $\hat H_{\rm int}$. In order to also flip the sign of the terms $\kappa_{\pm,i}^\dagger\kappa_{\pm,i+1}$, which can be seen as the hopping of a spin $m=0$, we multiply the staggered configuration by a phase factor $\pi\sum i_k$ where $i_k$ is the position indices of the $m=0$ states, as seen in Eq.\,(\ref{Eq: Staggering}).

We examine the accuracy of the descriptions presented above in Figure \ref{FigureSM_RQMBS}. The position of the N-QMBS in the energy spectrum is well reproduced. The  expectation value of $\hat n_0$ is in relatively good agreement, although clear deviations are observed. In particular the $p\hat S_x$ coupling gives rise to towers of N-QMBS with atypically low but finite expectation values $\langle\hat n_0\rangle$ while we strictly have $\langle\hat n_0\rangle=0$ with the Ising representation. Similarly, the Fock state description tends to overestimate the value of $\langle\hat n_0\rangle$. In order to have an indicator independent of an observable, we look at the weight $W_{I,F,SF}$ of each eigenstate within the three subspaces described above, namely ``Ising state" (corresponding to the $N_0=0$ sector), ``Fock state" (within the $N_\pm=0$ sectors and $\mathcal{H}_{\rm ex}$) and ``staggered Fock states" (within the $N_+=N_-$ sectors) . We observe that N-QMBS have remarkably large overlap, although not always reaching values close to $1$, in particular for the N-QMBS with the Fock state representation.

We perform this analysis for three system sizes $N=8\,,10$ and $12$. However, for such a limited range of system sizes it is challenging to make conclusive statements about the robustness of N-QMBS in the thermodynamic limit. A solution is provided by infinite-TEBD simulations which show robust oscillations (see Fig.\,\ref{Figure5}), and indicate that at least the N-QMBS close the $N_0=N$ Fock state survives in the thermodynamic limit.

\section{Thermal QMBS}\label{App: T-QMBS}

\begin{figure}
    \centering
    \includegraphics[width=\linewidth]{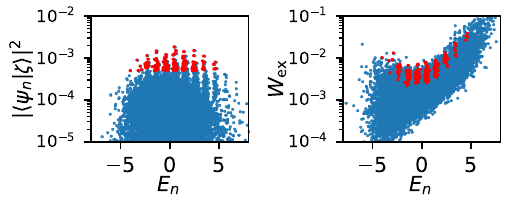}
    \caption{Left panel: Overlap of the eigenstates $\ket{\psi_n}$ with the coherent state $\ket{\mathbb{\zeta}}$ with ($n_0=0.5$,$\theta=4\pi/3$,$m=\eta=0$) taken on the UPO (same as in Fig.\,\ref{Figure3} (c) of the main text). Right panel: Weight of the eigenstates on the exchange symmetric energy sector $\mathcal{H}_{\rm ex}$. We highlight with red markers the states with an overlap $|\langle\psi_n\ket{\mathbb{\zeta}}|^2>5\times10^{-4}$. This threshold is arbitrary, but we observed that the states with the largest overlap are also the one showing the strongest quantum scars in their Husimi distribution. Such states are among those with largest $W_{\rm ex}$ (at a given energy), but they do not stand out from the rest of the spectrum. The small value of $W_{\rm ex}$ is the reason why T-QMBS does not prevent thermalization.}
    \label{FigureSM_Wex}
\end{figure}

We address  the relation between thermal behavior and quantum scarring. As mentioned in the main text, while T-QMBS show clear scars in their Husimi distribution, they do verify the ETH and carry high, volume-law entanglement. The Husimi distribution is defined at every point $\bm{\zeta}$ of the classical phase as the overlap with the SU(3) coherent states $\ket{\bm{\zeta}}$. The latter form an over-complete basis of the exchange symmetric manifold $\mathcal{H}_{\rm ex}$ with  dimension $\propto N^2$. Thus, with the Husimi distribution we only probe a very small fraction $\sim N^23^{-N}$ of the Hilbert space. We present in Fig.\,(\ref{FigureSM_Wex}) the weight $W_{\rm ex}$ of each eigenstate within $\mathcal{H}_{\rm ex}$. It is smoothly increasing with energy, which can be understood from the fact that the interaction energy is strictly positive for exchange symmetric states. Simultaneously, we provide the overlaps of the eigenstates with the coherent state $n_0=0.5,\theta=4\pi/3,m=\eta=0$ taken on the UPO at $E=0$ which shows towers of T-QMBS. We observe that the latter tends to be among the states with large $W_{\rm ex}$, but they do not constitute clear outliers. Furthermore, we also checked that for QMBS, and as well as for any other thermal states, $W_{\rm ex}$ appears to be exponentially decreasing in the range $N\in[6,12]$. Therefore, although the projection of T-QMBS onto $\mathcal{H}_{\rm ex}$ yields non-ergodic behavior by scarring, this still concerns only a small fraction of the many-body state. Hence the latter is thermal, and verifies the ETH featuring volume-law entanglement entropy.

\section{TDVP}\label{App: TDVP}

Below, we detail the calculation leading to 
the mean-field equations of motion and the quantum leakage.
\subsection{Equation of motion}
With the parametrization of Eq.\,(\ref{Eq: MF spinor}) the mean-field energy per spin is $\epsilon_s=\epsilon_{\rm int}+ps_x$ where 
\begin{subequations}
\begin{align}
    \epsilon_{\rm int}&=n_0(1-n_0)+\frac{m^2}{2},\\
    s_x &=\sqrt{2n_0}\left(\sqrt{n_+}\cos\phi_++\sqrt{n_-}\cos\phi_-\right)\,.
\end{align}
\end{subequations}
The Lagrangian for this system reads \cite{evrard2023quantum}
\begin{align}
	\mathcal{L}(n_0,\theta,m,\eta,...)&=\frac{i}{2}(\bm \zeta^{\dagger}\dot{\bm \zeta}-\textrm{c.c.})-\epsilon_s\,,\\
	&=-\frac{1-n_0}{2}\dot\theta-\frac{1}{2}m\dot\eta-\epsilon_s\,,
\end{align}
where we defined $\dot{\bm{\zeta}} := d\bm{\zeta}/dt$. The minimization of the action $\int dt\mathcal{L}$ yields
the Euler-Lagrange equations
\begin{align}
	\dot n_0 &=-2\frac{\partial \epsilon_s}{\partial\theta}\,,\hspace{0.5cm}
	\dot \theta =2\frac{\partial \epsilon_s}{\partial n_0}\,,\nonumber\\
	\dot \eta &= -2 \frac{\partial \epsilon_s}{\partial m}\,,\hspace{0.35cm} \dot m = 2 \frac{\partial \epsilon_s}{\partial\eta}\,.
\end{align}
After a straightforward algebra, we arrive at the equations of motion Eq.~(\ref{eq:eqnMotion}). In the calculation of the quantum leakage however, it is convenient to use the complex numbers $\zeta_m$ as variables instead of the population and phase. In that case we have
\begin{subequations}
\begin{align}
\epsilon_{\rm int} &= |\zeta_0|^2(|\zeta_+|^2+|\zeta_-|^2)+\frac{1}{2}\left(|\zeta_+|^2-|\zeta_-|^2\right)^2\\
 s_x&=\frac{p_x}{\sqrt{2}}\left(\zeta_+^*\zeta_0+\zeta_-^*\zeta_0+\rm{c.c.}\right)\,.
\end{align}
\end{subequations}
The Lagrangian in this set of variables can be written as
\begin{align}
	\mathcal{L}(\zeta_m,\zeta_m^*)=\frac{i}{2}\left(\sum_m \zeta_m^*\dot\zeta_m-\textrm{c.c.} \right)-\epsilon_s\,,
\end{align}
where $m=0,\pm$ resulting in the corresponding Euler-Lagrange equations
\begin{align}
        i\dot\zeta_m &=\frac{\partial \epsilon_s}{\partial \zeta_m^*}\,, \nonumber\label{Eq. DotZeta}\\
        i\frac{d}{dt}\begin{pmatrix}
		\zeta_+ \\\zeta_0\\\zeta_-
	\end{pmatrix} &=\begin{pmatrix}
	m & \frac{p_x}{\sqrt{2}} & 0\\
	\frac{p_x}{\sqrt{2}} & 1-2n_0 & \frac{p_x}{\sqrt{2}}\\
	0 & \frac{p_x}{\sqrt{2}} & -m\\
\end{pmatrix}\begin{pmatrix}
\zeta_+ \\\zeta_0\\\zeta_-
\end{pmatrix}.
\end{align}

\subsection{Quantum leakage}
The quantum leakage $\gamma(\bm{\zeta})$ is defined as the difference between the ``variational" derivative and the exact one given by Schrodinger equation:
\begin{align}
    \gamma^2(\bm{\zeta}) = \frac{1}{N}\left\Vert \ii\hat H\ket{\bm{\zeta}}+\left(\frac{d}{dt}\ket{\bm{\zeta}}\right)_{\rm mf}\right\Vert^2\,.\label{Eq: def leakageApp}
\end{align}
The variational derivative for our mean-field ansatz is
\begin{align}
    \left(\frac{d}{dt}\ket{\bm{\zeta}}\right)_{\rm mf}=\sum_{\substack{i_0=1..N\\m=0,\pm1}} \dot\zeta_m\ket{m}_{i_0}\bigotimes_{i\neq i_0}\ket{\bm{\zeta}}_i\,,\label{eq:D9}
\end{align}
due to the fact that $\ket{\bm{\zeta}}$ is a product state by definition.
Here $\dot\zeta_m$ is given by Eq.\,(\ref{Eq. DotZeta}) which we rewrite as
\begin{align}
&\ii\dot{\bm{\zeta}}=\nabla_{\bm{\zeta^*}}\epsilon_{\rm int}+p\nabla_{\bm{\zeta^*}}s_x\label{eq:D10}
\end{align}
By substituting Eq.~\eqref{eq:D10} into Eq.~\eqref{eq:D9}, the mean-field derivative can be written as
\begin{align}
	\left(\frac{d}{dt}\ket{\bm{\zeta}}\right)_{\rm mf} = \hat D\ket{\bm{\zeta}}-ip\hat S_x\ket{\bm{\zeta}}\,,
\end{align}
where we introduced the operator
\begin{align}
    \hat D\ket{\bm{\zeta}} = -\ii\sum_{i_0,m} \nabla_{\bm{\zeta^*}}\epsilon_{\rm int}\ket{m}_{i_0}\bigotimes_{i\neq i_0}\ket{\bm{\zeta}}_i\,.
\end{align}
We see that term single-particle term $\hat S_x$ contributes to the same way both for the exact and mean-field derivatives and thus cancel out in the expression of the quantum leakage. Therefore this term can be omitted in the rest of the calculation, and for simplicity let us restate $\dot{\bm{\zeta}} = -\ii\nabla_{\bm{\zeta^*}}\epsilon_{\rm int}$ leading to $\ket{\dot{\bm{\zeta}}}=\hat D\ket{\bm{\zeta}}$.
An important property of Eq.\,(\ref{Eq: def leakageApp}) is that it is gauge-dependent~\cite{Michailidis2020,Ho2019}. In order to enforce that the global phase of the variational state is the same as the one obtained from the Schr\"odinger equation, we change the gauge $\ket{\bm{\zeta}}\to\ee^{\ii\phi}\ket{\bm{\zeta}}$ with 
\begin{align}
\dot\phi = -\langle\hat H_{\rm int}\rangle+\ii\bra{\bm{\zeta}}\dot{\bm{\zeta}}\rangle\,,
\end{align}
We thus have
\begin{align}
    \gamma^2 &= \frac{1}{N}\left\Vert \ii \left(\hat H_{\rm int}-\langle\hat H_{\rm int}\rangle\right)\ket{\bm{\zeta}}+\ket{\dot{\bm{\zeta}}}-\bra{\bm{\zeta}}\dot{\bm{\zeta}}\rangle\ket{\bm{\zeta}}\right\Vert^2\,.
\end{align}
By developing the norm, we can write $\gamma^2 = \frac{1}{N}(C_1+C_2+C_3)$ where
\begin{subequations}
\begin{align}
    C_1&=\bra{\dot{\bm{\zeta}}}\dot{\bm{\zeta}}\rangle-\vert\bra{\bm{\zeta}}\dot{\bm{\zeta}}\rangle\vert^2\,,\\
	C_2&=2\mathrm{Im}\left(\bra{\bm{\zeta}}\hat H_{\rm int}\ket{\dot{\bm{\zeta}}}-\langle\hat H_{\rm int}\rangle\bra{\bm{\zeta}}\dot{\bm{\zeta}}\rangle\right)\,,\label{eq:D15b}\\
	C_3&=\langle\hat H_{\rm int}^2\rangle-\langle\hat H_{\rm int}\rangle^2\,.
\end{align}
\end{subequations}
Note that the disconnected correlators cancel thanks to the proper gauge choice such that we are left with connected correlators only, which are of order $C_j\sim N$ for mean field states (as we explicitly show next). This leads to $\gamma\sim 1$ in general. An exception occurs in the all-to-all interacting Hamiltonian where the leading contributions of the $C_j$ exactly cancel and $\gamma\sim 1/\sqrt{N}$, as shown in Sec.~\ref{App: TDVP all-to-all}.

\paragraph{Computation of $C_1$.}

The different contributions correspond to
\begin{subequations}
\begin{align}
	&\bra{\bm{\zeta}}\dot{\bm{\zeta}}\rangle = N \bm{\zeta}^{\dagger} \dot{\bm{\zeta}} = N\sum_m \zeta_m^* \dot\zeta_m\,,\label{eq:D16a}\\
&\bra{\dot{\bm{\zeta}}}\dot{\bm{\zeta}}\rangle = N(N-1)\big| \bm{\zeta}^{\dagger} \dot{\bm{\zeta}} \big|^2+N|\dot{\bm{\zeta}}|^2\,.	
\end{align}
\end{subequations}
This leads to 
\begin{align}
    C_1= N\bigg( \big|\dot{\bm{\zeta}}\big|^2 -\big| \bm{\zeta}^{\dagger} \dot{\bm{\zeta}} \big|^2 \bigg)\,,
\end{align}
with 
\begin{align}
	\ii \begin{pmatrix}
		\dot \zeta_+ \\ \dot \zeta_0\\ \dot \zeta_-
	\end{pmatrix}=\nabla_{\bm{\zeta}^*}\epsilon_{\rm int}=\begin{pmatrix}
		m \hspace{1mm} \zeta_+\\
		(1-2n_0)\hspace{1mm}\zeta_0\\
		-m\hspace{1mm}\zeta_-\\
	\end{pmatrix}\,,
\end{align}
we arrive at
\begin{align}
	\frac{C_1}{N}=n_0(1-n_0)(1-2n_0)^2+m^2(1-3n_0+4n_0^2)-m^4\,.\label{Eq: C1}
\end{align}
\paragraph{Computation of $C_2$.}~Let us first focus on the second term in Eq.~\eqref{eq:D15b}, 
\begin{subequations}
\begin{align}
	&\hat H_{\rm int}=\sum_{i=1}^{N}\hat{h}_{i,i+1}\,,\\
 &\langle\hat H_{\rm int}\rangle=N\langle\hat h_{12}\rangle=N\epsilon_{\rm int}\,,
\end{align}
\end{subequations}
where we have used the translation symmetry. This leads to $\langle\hat H_{\rm int}\rangle\bra{\bm{\zeta}}\dot{\bm{\zeta}}\rangle = N^2 \epsilon_{\rm int} \bm{\zeta}^{\dagger} \dot{\bm{\zeta}}$. 
\begin{align}
	\bra{\bm{\zeta}}\hat H_{\rm int}\ket{\dot{\bm{\zeta}}}=&\sum_{i}\sum_{j_0,m}\bra{\bm{\zeta}}\hat{h}_{i,i+1}\dot\zeta_m\ket{m}_{j_0}\bigotimes_{j\neq j_0}\ket{\bm{\zeta}}_{j}\,,\nonumber\\
	=&N(N-2)\big\langle\hat h_{12}\big\rangle \bm{\zeta}^{\dagger} \dot{\bm{\zeta}} \notag\\
&+ N \sum_m\sum_{j_0=1,2}\dot\zeta_m \prescript{}{1}{\bra{\bm{\zeta}}}\prescript{}{2}{\bra{\bm{\zeta}}}\hat h_{12}\ket{m}_{j_0}\ket{\bm{\zeta}}_{j\neq j_0}\,,\notag\\
=&N(N-2)\epsilon_{\rm int}\bm{\zeta}^*\cdot\dot{\bm{\zeta}}+N\nabla_{\bm{\zeta}}\epsilon_{\rm int} \dot{\bm{\zeta}}\,.
\end{align}
We then use $\nabla_{\bm{\zeta}}\epsilon_{\rm int}=-\ii\bm{\zeta}^\dagger$ and arrive at
\begin{align}
\frac{C_2}{N}&=2\mathrm{Im}\left[-2\epsilon_{\rm int} \bm{\zeta}^{\dagger} \dot{\bm{\zeta}} - 2\ii N \big|\dot{\bm{\zeta}}\big|^2 \right]\,.
\end{align}
Finally, noticing that $\epsilon_{\rm int}$ is a quadratic form, we have $2\epsilon_{\rm int}=\bm{\zeta}\nabla_{\bm{\zeta}}\epsilon_{\rm int}=-\ii\dot{\bm{\zeta}}^{\dagger}\bm{\zeta}$, leading to
\begin{align}
	\frac{C_2}{N}&=2\left(\big|\bm{\zeta}^{\dagger} \dot{\bm{\zeta}}\big|^2-\big|\dot{\bm{\zeta}}\big|^2 \right)=-2\frac{C_1}{N}\,.\label{Eq: C2}
\end{align}

\paragraph{Computation of $C_3$.}~Using the translational symmetry, we write
\begin{align}
	\big\langle \hat H_{\rm int}^2 \big\rangle &=\bigg\langle \sum_{i,j}\hat h_{i,i+1}\hat h_{j,j+1}\bigg \rangle\,, \\
	&=N\big \langle \hat h_{12}^2 \big \rangle + 2N \big \langle \hat h_{12}\hat h_{23} \big \rangle +  N(N-3)\big \langle \hat h_{12}\big \rangle^2.\notag
\end{align} 
These terms originate from the situations $i=j$; $i=j+1$ or $i=j-1$; and $i \neq j$, respectively. We straightforwardly compute
\begin{subequations}
\begin{align}
    &\big\langle \hat h_{12} \big \rangle=n_0(1-n_0)+\frac{m^2}{2}\,,\\
    &\big\langle\hat h_{12}^2 \big\rangle=\frac{1}{4}\big(1-n_0^2\big)\,,\\
    &\big\langle \hat h_{12}\hat h_{23}\big \rangle=\frac{1}{4}[n_0(1-n_0)+m^2(1+n_0)]\,,
    \end{align}
\end{subequations}
giving rise to
\begin{align}
    \frac{C_3}{N} =& \frac{1}{2}n_0(1-n_0)-3n_0^2(1-n_0)^2+\frac{1}{4}\big(1-n_0^2\big)\nonumber\\
    +&\frac{m^2}{2}\big(1-5n_0+6n_0^2\big)-\frac{3m^4}{4}\,.\label{Eq: C3}
\end{align}
Finally, using Eqs.\,(\ref{Eq: C1}), (\ref{Eq: C2}), (\ref{Eq: C3}), the leakage takes the form $\gamma^2=\frac{1}{N}(C_3-C_1)$ and after some rearrangements we obtain Eq.~\eqref{Eq. leakage explicit}. 

\subsection{All-to-all interaction}\label{App: TDVP all-to-all}
Let us briefly discuss the case of infinite-range interactions which was extensively studied in \cite{evrard2023quantum}. To obtain an extensive energy with the same mean-field limit as for the nearest neighbor interaction, the spin-spin interaction strength is rescaled by $1/N$. Then the Hamiltonian reads
\begin{align}
    \hat H_{\mathrm{int},\infty} &= \frac{1}{2N}\sum_{i\neq j=1}^{N}\Big(\hat \kappa_{+,i}^\dagger\hat \kappa_{+,j}+\hat \kappa_{-,i}^\dagger\hat \kappa_{-,j}+
\mathrm{H.c.}\nonumber\\
&+\hat s_{z,i}\hat s_{z,j}\Big)\,,\label{Eq: Hint all-to-all}\\
&=\frac{1}{2N}\left(2N_0(N_++N_-)+(N_+-N_-)^2+N_0-N\right)\,.\notag
\end{align}
Since the mean-field energy and equations of motion are preserved, the terms $C_1$ and $C_2=-2C_1$ in the expression of the quantum leakage are the same for the all-to-all interacting Hamiltonian. On the other hand, the term $C_3$ is modified as
\begin{align}
    \frac{C_3}{N}=&n_0(1-n_0)(1-2n_0)^2+m^2\big(1-3n_0+4n_0^2\big)-m^4\nonumber\\
    &+\frac{1}{N}\bigg(\frac{5}{2}m^4-10m^2n_0^2+8m^2n_0-3m^2+10n_0^4\nonumber\\&-22n_0^3+\frac{31}{2}n_0^2-4n_0+\frac{1}{2}\bigg)+\mathcal{O}\bigg(\frac{1}{N^2}\bigg)\,.
\end{align}
The first term is equal to $C_1/N$ which exactly cancels $(C_1+C_2)/N$. The quantum leakage thus scales as $\gamma\sim1/\sqrt{N}$ attesting the validity of the mean-field description in the thermodynamic limit.

\section{Unstable periodic orbit with $E=0$}\label{App: Pendulum}
In the manifold defined by $m=\eta=0$ the equations of motion for $n_0$ and $\theta$ take the following simple form
\begin{align}
    \dot n_0=& 2p\sqrt{n_0(1-n_0)}\sin\left(\frac{\theta}{2}\right)\,,\\
    \dot \theta=& 2(1-2n_0) +2p\frac{1-2n_0}{\sqrt{n_0(1-n_0)}}\cos\left(\frac{\theta}{2}\right)\,,
\end{align}
while the mean-field energy per spin reads
\begin{align}
    \epsilon_s&=n_0(1-n_0)+2p\sqrt{n_0(1-n_0)}\cos\left(\frac{\theta}{2}\right).
\end{align}
Let us introduce the new variable $\alpha=4\arccos(\sqrt{n_0})$ such that $n_0=\cos^2(\alpha/4)$ (prefactor $4$ will become evident). We obtain the following equations when the new variable is substituted into $\epsilon_s$,
\begin{eqnarray}
  \epsilon_s &=& \sin^2\left(\frac{\alpha}{4}\right)\cos^2\left(\frac{\alpha}{4}\right) \notag \\
  &+&2p\left|\sin\left(\frac{\alpha}{4}\right)\cos\left(\frac{\alpha}{4}\right)\right|\cos\left(\frac{\theta}{2}\right),\label{eq:F4}
\end{eqnarray}
and into the equation of motion for $n_0$,
\begin{align}
    -\frac{1}{2}\sin\left(\frac{\alpha}{4}\right)\cos\left(\frac{\alpha}{4}\right)\dot \alpha = 2p\left|\sin\left(\frac{\alpha}{4}\right)\cos\left(\frac{\alpha}{4}\right)\right|\sin\left(\frac{\theta}{2}\right).\label{eq:F5}
\end{align}
By combining Eqs.~\eqref{eq:F4} and~\eqref{eq:F5} we eliminate $\theta$. In the case of $\epsilon_s=0$, we arrive at Eq.~\eqref{Eq: Pendulum} which describe the motion of a pendulum as presented in the main text.

\bibliographystyle{apsrev4-1}

\begin{thebibliography}{60}
\bibitem [{Foo({\natexlab{c}})}]{FootNotes_NormalizationGamma}%
  \BibitemOpen
  \href@noop {} {}\bibinfo {howpublished} {Different normalization of $\gamma$
  can be found in the literature. For the one we used here, the general
  expectation is $\gamma\sim1$. Note that for all-to-all interaction, as shown
  in the appendix, we have $\gamma\sim N^{-1/2}$, signaling that the mean-field
  description becomes exact in the limit $N\to\infty$.}
  ({\natexlab{c}})\BibitemShut {NoStop}%
\end{thebibliography}

%

\end{document}